\documentclass[aip,reprint,jcp,graphicx,amssymb,
]{revtex4-1}
\pdfoutput=1
\usepackage{graphicx}
\graphicspath{{Figures/}}
\usepackage{array}
\usepackage{bm}
\usepackage{rotating}
\usepackage{dcolumn}
\usepackage{outlines}
\usepackage{latexsym}
\usepackage{amsfonts}
\usepackage{amssymb}
\usepackage[usenames,dvipsnames,table]{xcolor}
\usepackage{tikz}
\newcommand*\circled[1]{\tikz[baseline=(char.base)]{
            \node[shape=circle,draw,inner sep=0.5pt] (char) {#1};}}

\newcommand{\reda}[1]{\textcolor{red}{#1}}
\newcommand{\green}[1]{\textcolor{green}{#1}}
\newcommand{\blue}[1]{\textcolor{blue}{#1}}
\usepackage[version=4]{mhchem} 

\begin{document}


\title{Isotope-specific reactions of acetonitrile (\ce{CH3CN}) with trapped, translationally cold \ce{CCl+}}
\author{O. A. Krohn}
\email{olivia.krohn@colorado.edu}
\author{K. J. Catani}
\author{J. Greenberg}
\affiliation{Department of Physics, University of Colorado, Boulder, Colorado, USA}
\affiliation{JILA, National Institute of Standards and Technology and the University of Colorado, Boulder, Colorado, USA}
\author{S. P. Sundar}
\author{G. da Silva}
\affiliation{Department of Chemical Engineering, The University of Melbourne, Parkville 3010, Victoria, Australia}
\author{H. J. Lewandowski}
\affiliation{Department of Physics, University of Colorado, Boulder, Colorado, USA}
\affiliation{JILA, National Institute of Standards and Technology and the University of Colorado, Boulder, Colorado, USA}

\date{\today}

\begin{abstract} 
The gas-phase reaction of \ce{CCl^+} with acetonitrile (\ce{CH3CN}) is studied using a linear Paul ion trap coupled to a time-of-flight mass spectrometer. This work builds on a previous study of the reaction of \ce{CCl+} with acetylene\cite{Catani2020} and further explores the reactivity of \ce{CCl+} with organic neutral molecules. Both of the reactant species are relevant in observations and models of chemistry in the interstellar medium (ISM). Nitriles, in particular, are noted for their relevance in prebiotic chemistry, such as is found in the atmosphere of Titan, one of Saturn's moons. This work represents one of the first studied reactions of a halogenated carbocation with a nitrile, and the first exploration of \ce{CCl+} with a nitrile. Reactant isotopologues are used to unambiguously assign ionic primary products from this reaction: \ce{HNCCl+} and \ce{C2H3+}. Branching ratios are measured and both primary products are determined to be equally probable. Quantum chemical and statistical reaction rate theory calculations illuminate pertinent information for interpreting the reaction data, including: reaction thermodynamics, a potential energy surface for the reaction, as well as rate constants and branching ratios for the observed products. In particular, the reaction products and potential energy surface stimulate questions regarding the strength and role of the nitrile functional group, which can be further explored with more reactions of this class.
\end{abstract}

\maketitle

\section{Introduction}

Nitriles and nitrogen-containing compounds play a prominent role in the chemical reactions thought to take place in the interstellar medium (ISM). These molecules permeate space: from small cyanides such as HCN and DCN found in the Orion Nebula\cite{Snyder1971,JEFFERTS1973} to larger molecules such as benzonitrile, whose initial discovery in the ISM was relatively recent.\cite{science_benzo} Nitriles, defined by their \ce{C#N} functional group, are of particular interest as pre-biotic molecules and potential precursors of amino acids. Several nitriles have been identified in the atmosphere of Titan using the Ion Neutral Mass Spectrometer on the Cassini spacecraft, and are believed to be important in tholin formation,\cite{Waite2007} as well as astrobiology.\cite{Ali2015} 

Acetonitrile (\ce{CH3CN}; the neutral reactant in this study) has been found abundantly throughout many regions of space since its initial identification in the ISM in 1971.\cite{Solomon1971} It has been observed in cold dark clouds,\cite{Matthews1983} low-mass protostars,\cite{Codella2009, Bisschop2008} and is considered an indicator of the presence of hot cores.\cite{Cazaux_2003, bisschop2007} \ce{CH3CN} has also been discovered in dust from comet Halley,\cite{Kissel1987} Hale-Bopp (C/1995 O1)\cite{Biver1997} and, more recently, at the surface of comet 67P/Churyumov-Gerasimenko.\cite{Morse2019} These cometary identifications can yield critical glimpses into the past conditions and evolutionary history of the Milky Way.
Deuterated variants \ce{CD3CN} and \ce{CDH2CN} have been identified in hot cores and star-formation regions,\cite{Gerin1992} and the presence of isotopologues of \ce{CH3CN} are used to study relative populations of hydrogen and deuterium in some regions of the ISM.\cite{Belloche2016deut} 

Halogen-containing compounds have also been identified in the ISM, but their role and evolution are less well understood. In particular, chlorine-containing compounds have been found in the ISM in several small molecules (NaCl, AlCl, KCl, HCl),\cite{McGuire2018} as well as in \ce{CH3Cl}\cite{Fayolle2017} and \ce{H2Cl+}.\cite{Lis2010,Neufeld_2012} The only halogenated carbocation to be observed thus far in the ISM is \ce{CF+},\cite{McGuire2018} whereas \ce{CCl+} has been predicted to occur, although only in low abundances.\cite{Neufeld2009} \ce{CCl+} can be produced from reactions of \ce{C+ + HCl},\cite{Glosik1993} and once formed, has been assumed to be predominantly nonreactive. Specifically, \ce{CCl+} has been shown to not react with HCN (or \ce{CO2}, CO, \ce{O2}, \ce{H2O}, \ce{CH4}, \ce{H2}). However, it has been shown to react with \ce{NH3} and \ce{H2CO}.\cite{Blake1986} Recent work from our group demonstrated \ce{CCl^+} reacts with acetylene (\ce{C2H2}), producing small fundamental carbocations after losing neutral Cl or \ce{HCl}.\cite{Catani2020} Despite this, much remains unknown about the role of halogenated carbocations; it is possible that they have a hitherto underestimated role in astronomical chemistry.

In contrast to \ce{CCl+}, laboratory reactions of nitriles have been much more widely studied. Ion cyclotron resonance (ICR) spectrometry has been used to measure reactions with HCN and carbocations,\cite{Anicich1986} while other ion trap experiments have investigated reactions of \ce{CH3CN} with multiple carbocations.\cite{Blair1973} Selected-ion flow-tube mass spectrometry (SIFT) experiments demonstrated reactivity of \ce{CH3CN} with \ce{O+}, \ce{H+}, \ce{D+}, \ce{HeD+}, and \ce{HeH+},\cite{SmithSIFT1992} as well as with \ce{C2H4+},\cite{Petrank1992} and \ce{C2H2+}.\cite{IRAQI1990} However, very few measurements have reported reactions of halogenated carbocations with any nitrile. The only reported reaction of this type is the reaction of \ce{CF3+} with \ce{CH3CN} and benzonitrile, both of which were shown to produce only the adduct.\cite{Tsuji1995} The reactions of \ce{CF3+} were executed in a higher pressure regime than that of the current experiment, where reactive intermediates are unable to be stabilized through collisions with background gas. The reactivity of halogenated carbocations with nitriles is in need of further exploration, particularly in a cold, low-pressure environment. This work seeks to understand more about this reaction class by studying the reaction of \ce{CCl^+ + CH3CN} in this regime.

The cold, low-pressure environment provided by using a linear Paul ion trap (LIT) is excellent for elucidating ion-neutral chemical reactions.\cite{Heazlewood2019,Toscano2020} This experimental setup affords a significant amount of control, including the manipulation of collisional energy,\cite{Puri2019,okadaRSI2017} nuclear spin,\cite{Kilaj2018} and the measurement of isomer,\cite{Chang13, schmidgreenberg2020} isotope,\cite{Lavert-Ofir2014,Petralia2020} and quantum state\cite{SchmidMOL2019,GreenbergPRA2018} dependencies. Ions of interest are co-trapped and sympathetically cooled with laser-cooled \ce{Ca+}, forming a mixed species Coulomb crystal, achieving translationally cold, trapped ions. Furthermore, the addition of a time-of-flight mass spectrometer (TOF-MS) provides detection of ionic reactants and products with high mass resolution -- a powerful tool for probing reaction products and kinetics.

The reaction of sympathetically cooled \ce{CCl+} with \ce{CH3CN} is studied using our LIT TOF-MS. This work seeks to illuminate the role and reactivity of these novel species in the gas phase under experimental conditions that are approximate to that of the ISM and planetary atmospheres. The primary products are found to be \ce{C2H3+} and \ce{HNCCl+}, which are unambiguously assigned through the use of isotope substitutions. Computational modeling also supports these product assignments, suggesting a reaction pathway requiring cleavage of the \ce{C#N} bond of \ce{CH3CN} in order to form the observed products. Furthermore, the study of \ce{CCl^+ + CH3CN} signifies an initial investigation in reactions of halogenated carbocations with nitriles. 

\section{Methods}
\subsection{Experimental Methods}
\begin{figure}[htb]
    \centering
    \includegraphics[ width=8.5cm]{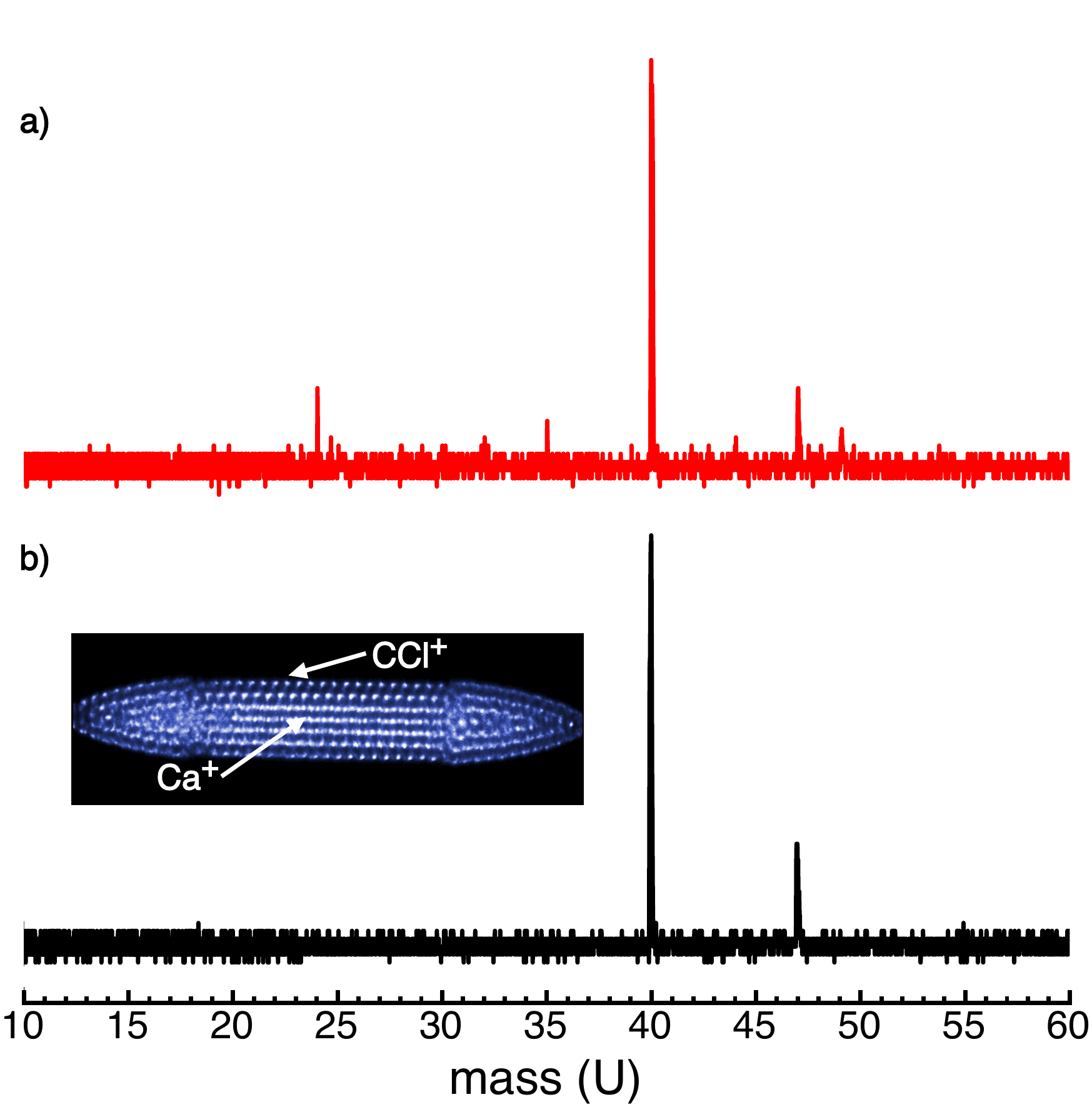}
    \caption {\textbf{a)} TOF traces demonstrating before and \textbf{b)} after cleaning using secular excitations. After cleaning, only \ce{Ca+} ($m/z$ 40, $m/z$ 42, and $m/z$ 44) and \ce{CCl+} ($m/z$ 47) remain in quantities greater than $\sim5$ ions. Also included on the left is a false-color CCD image of fluorescing \ce{Ca+} ions, the resulting Coulomb crystal is deformed primarily in the center section by the heavier \ce{CCl+} ions. The crystal appears truncated because it expands slightly beyond the CCD camera frame.}
    \label{fig:crystal}
\end{figure}

Reaction data were collected using a LIT radially coupled to a TOF-MS. Detailed descriptions of the apparatus have been outlined previously,\cite{SchmidRSI2017,Catani2020} and only a brief summary focusing on the specific details relevant to the current experiment will be given here. \ce{CCl+} was produced using tetrachloroethylene (TCE, \ce{C2Cl4}) seeded in a pulsed supersonic expansion of rare atomic gas (1.4\% \ce{C2Cl4} in $\sim$1000\,Torr He). The skimmed molecular beam was overlapped with a focused beam (216\,nm) from a pulsed dye laser (LIOPTEC LiopStar; 10\,ns pulse, 100\,$\mu$J/pulse) in the center of the trap. Non-resonant multiphoton ionization of TCE resulted in several fragments, including \ce{C^35Cl+}, \ce{C^37Cl+}, \ce{^35Cl+}, \ce{^37Cl+}, \ce{C2+}, and small amounts of \ce{C2 ^35Cl+} (hereafter, the more abundant isotope \ce{^35Cl} will be referred to as simply \ce{Cl}, while \ce{^37Cl} will be specified when appropriate). Unwanted ions were ejected from the trap by sweeping over resonance frequencies of the specific mass-to-charge ratio ($m/z$) of undesired ions.\cite{Roth2007} This provided a clean sample of either \ce{CCl+} or \ce{C^37Cl+} with minimal impurities, as demonstrated in Fig. \ref{fig:crystal}.

After removing unwanted ionization products from the trap, \ce{Ca+} was loaded by non-resonantly photoionizing an effusive beam of calcium using the third harmonic of an Nd:YAG (Minilite, 10\,Hz, $\sim7$\,mJ/pulse). The resulting \ce{Ca+} ions were Doppler laser cooled by two external cavity diode lasers, forming a Coulomb crystal structure, which sympathetically cooled the co-trapped \ce{CCl+} ions via the Coulomb interaction. \ce{Ca+} ion fluorescence was collected using a microscope objective and focused onto an intensified CCD camera located above the trap, allowing for qualitative visual monitoring of the experiment. The heavier ``dark" \ce{CCl+} ions arrange themselves in outer shells around the \ce{Ca+} ions, deforming the fluorescing Coulomb crystal as seen in Fig. \ref{fig:crystal}b. A typical experiment utilized 150-250 \ce{CCl+} ions trapped with $\sim1000$ \ce{Ca+} ions, all of which were translationally cold ($\sim10$\,K).

After \ce{CCl+} and \ce{Ca+} ions were loaded, neutral \ce{CH3CN} (9-10\% \ce{CH3CN} or \ce{CD3CN} in \ce{N2}) was leaked into the vacuum chamber ($2\times10^{-9}$\,Torr gas pressure at 300\,K) for a set duration of time using a pulsed leak-valve scheme.\cite{Jiao1996,SchmidMOL2019} The measurements of gas partial pressures in the chamber were recorded using a Bayard-Alpert hot cathode ionization gauge. The opening of the leak valve (LV) defined the zero-time point; the LV remained open for 0, 10, 30, 60, 90, 120, 150, 180, 210, 240, or 330\,s before ejecting the ions into the TOF-MS. This process was repeated about 10 times for every time step and measured ion numbers from each mass were averaged over each time step. The average number of reactant and product ions were then normalized by the initial \ce{CCl+} numbers and plotted against time, forming a reaction curve. These reaction curves were then used to determine the relevant rate constants. Reaction curves were collected in the same manner for isotopologues \ce{C^37Cl+} and \ce{CD3CN}, such that all four possible combinations of isotopologues were used. The chemical formula of each mass peak was confirmed by examining the shift in mass spectra as a result of isotopologue substitution (see section \ref{ExpRes}). In addition, all of the ionic species were tracked via TOF-MS traces. The total number of ions were compared for each time point to ensure that the numbers were constant throughout the experiment; this ruled out systematic losses of ions from the trap. Figures illustrating conservation of charge over each reaction are given with more context in the supplementary material. 

\subsection{Computational Methods}
Several theoretical methods were used to explore the potential energy surface for the reaction of \ce{CCl+ + CH3CN}. In a previous study, the M06-2X/aug-cc-pVTZ level of theory was found to produce accurate geometries and energies for small nitrogen- and chlorine-containing compounds,\cite{Castet2012} and was therefore chosen to determine possible stationary points. Scans over bond lengths, angles, and dihedrals allowed identification of minima and saddle points. Transition states were verified by visually inspecting the single imaginary frequency and also by using intrinsic reaction coordinate (IRC) analysis. The geometries of the reactants, products, intermediate states, and transition states were then used as starting points for calculations at the MP2/aug-cc-pVTZ level of theory. Zero point energy (ZPE) corrections from calculated harmonic vibrational frequencies (MP2/aug-cc-pVTZ) were added to CCSD(T)/CBS single point energies [CCSD(T)/CBS//MP2/aug-cc-pVTZ nomenclature is used in the subsequent discussions]. Additional higher order calculations were carried out at the CCSD(T)/CBS//CCSD/aug-cc-pVTZ level of theory for reactants and predicted products to provide accurate energetics for the thermodynamic limits of the reaction within 0.04\,eV. 
Even though \ce{^37Cl} and D isotope substitutions were used experimentally to determine the chemical formulas of the products, calculations accommodating these substitutions are outside the scope of this work. Density functional theory (DFT) calculations and relaxed potential energy surface scans were done using Gaussian 16,\cite{g16} while the higher order MP2 and CCSD computations were done using Psi4 v1.3.2.\cite{psi4}

Statistical reaction rate theory calculations were performed to simulate the kinetics of the \ce{CCl+ + CH3CN} reaction. These calculations were carried out using a custom version of the MultiWell2020 suite of programs,\cite{multiwell, barker2001, barker2009} modified to treat bath-gas collisions using the Langevin model. Simulations followed a general approach that we have used extensively to investigate ion reaction dynamics in a diverse range of instruments, including ion trap,\cite{GdS2012} tandem,\cite{Catani2017} and ion mobility\cite{Bull2018} mass spectrometers.
Electronic energies, vibrational frequencies, and moments of inertia were from the CCSD(T)/CBS//MP2/aug-cc-pVTZ model chemistry calculations. Microscopic rate constants were calculated via Rice-Ramsperger-Kassel-Marcus (RRKM) theory, on the basis of rigid-rotor harmonic-oscillator sums and densities of state. For barrierless ion-molecule reactions, association rate coefficients were set at the ADO theory value, with the restricted Gorin model\cite{gorin} then applied to fit an effective transition state structure. Energy grained master equation simulations were performed in order to predict the \ce{CCl+ + CH3CN} reaction products. These calculations featured energy grains of 10\,cm$^{-1}$ and a single exponential down collisional energy transfer model, with the average energy in deactivating collisions set at 200\,cm$^{-1}$.\cite{DEdown} Simulations comprised 10$^{10}$ trajectories, and in each case a reaction was predicted to be complete within less than the time required for one bath-gas collision (i.e., effectively collisionless). Simulations were performed at a pressure of $2\times10^{-9}$\,Torr \ce{N2}, with temperature varied between 40 and 400 K in order to examine predicted rates from atmospheric down to astrochemically relevant conditions.

\section{Results \& Discussion}
For the sake of clarity, the reaction thermodynamics will be discussed with the concluded chemical formula assignments in Section \ref{therm}, followed by experimental support in Section \ref{ExpRes}. Finally, in Section \ref{ModelPES} the modeled potential energy surface, branching ratios, and rate constants of the reaction are discussed. 
\subsection{Reaction thermodynamics}\label{therm}
\begin{figure}[!]
    \centering
    \includegraphics[width=8.5cm]{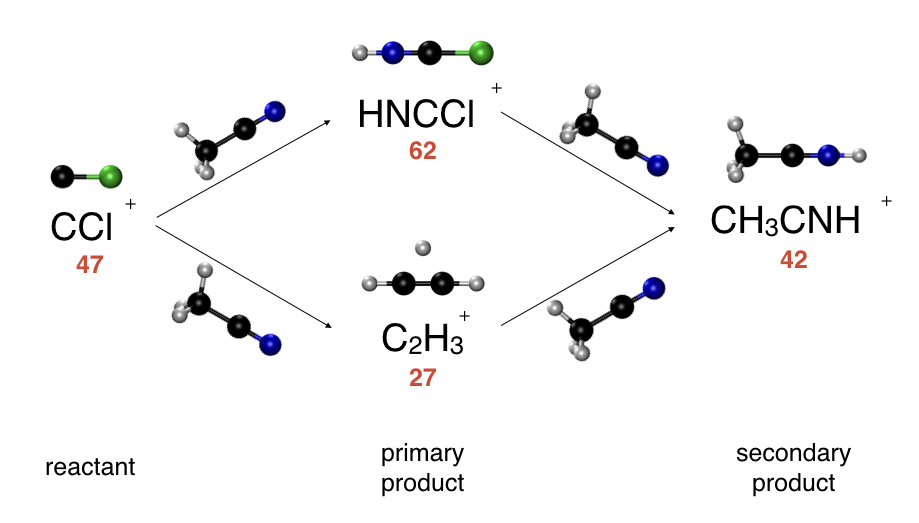}
    \caption{Reaction model for \ce{CCl+}+ \ce{CH3CN}, noting the reaction order and identity of ions. Each arrow represents a reaction with a neutral \ce{CH3CN} molecule. Red number below the molecule denotes $m/z$ ratio. The molecular ions are depicted above, with black indicating carbon, blue for nitrogen, white for hydrogen, and green for chlorine.}
    \label{fig:model}
\end{figure} 
 Overall, the reaction of \ce{CCl+ + CH3CN} forms the primary ionic products \ce{C2H3+} and \ce{HNCCl+}, which proceed to react with excess \ce{CH3CN} to form the secondary product protonated acetonitrile (\ce{CH3CNH+}). This model is illustrated in Fig. \ref{fig:model}.

Neutral \ce{CH3CN} was introduced into the vacuum chamber as a room temperature gas (300\,K). Therefore, when reacting with translationally cold \ce{CCl+} ($\sim10$\,K), the calculated collision energy for the reaction is $\sim15$\,meV (160\,K). This provides a narrow upper limit to the reaction energetics. The observed products are all significantly exothermic and well below the upper limit provided by the calculated collision energy, as shown by Equations \ref{p1}-\ref{s4} [CCSD(T)/CBS//CCSD/aug-cc-pVTZ; accurate within 0.04\,eV].

Primary products:
\begin{equation}\label{p1}
\begin{split}
\ce{CCl+ + CH3CN -> C2H3+ + NCCl} \\
\Delta E = -1.17\,\text{eV}
\end{split}
\end{equation}
\begin{equation}\label{p2}
\begin{split}
\ce{CCl+ + CH3CN -> HNCCl+ + C2H2}\\
\Delta E = -2.09\,\text{eV}
\end{split}
\end{equation}

Secondary products:
\begin{equation}\label{s3}
\begin{split}
\ce{C2H3+ + CH3CN -> CH3CNH+ + C2H2} \\
\Delta E = -1.41\,\text{eV}
\end{split}
\end{equation}
\begin{equation}\label{s4}
\begin{split}
\ce{HNCCl+ + CH3CN -> CH3CNH+ + NCCl} \\
\Delta E = -0.48\,\text{eV}
\end{split}
\end{equation}

These calculated limits assume the lowest energy isomers. For example, in Eqns. \ref{p2} and \ref{s3}, the \ce{C2H3+} energy refers to that of the non-classical ``bridge'' isomer (see Fig. \ref{fig:model} or PRD2 in Fig. \ref{fig:CCl+CH3CN_PES}). This non-classical isomer is where the third H hovers between the two carbons, as opposed to the ``classical'' or ``Y'' structure (\ce{H2C2H+}, see PRD3).
Other possible isomeric products are discussed in Section \ref{ModelPES}. 

\subsection{Reaction measurements}\label{ExpRes}
\begin{figure}[h]
    \centering
    \includegraphics[width=8.5cm]{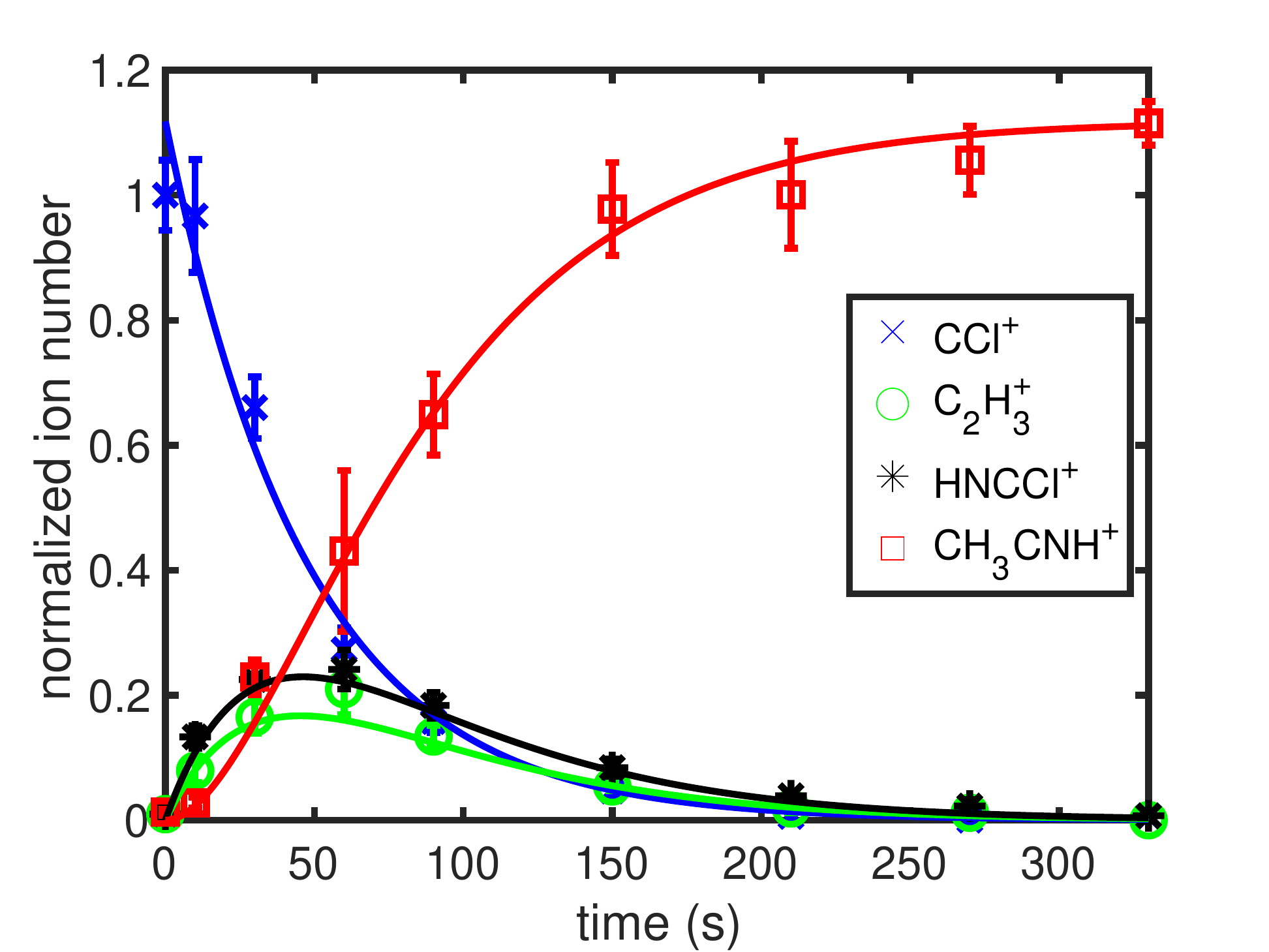}
    \caption{Rate reaction data (points) and fits (curves) for pseudo-first order reaction of \ce{CCl+ + CH3CN}. \ce{CCl+} (blue\,\blue{$\bm{\times}$}) reacts with excess \ce{CH3CN} resulting in first order products  \ce{C2H3+} (green\,\green{$\bm{\circ}$}) and \ce{HNCCl+} (black\,$\bm{\ast}$). Each of these primary products then reacts with  excess \ce{CH3CN} to form \ce{CH3CNH+} (red\,\reda{$\bm{\Box}$}). }
    \label{fig:reactioncurve}
\end{figure}

Curves that are produced from the reaction of \ce{CCl+ + CH3CN} are shown in Fig. \ref{fig:reactioncurve}. Here, \ce{CCl+} ($m/z$ 47; blue) reacts to form two primary products: \ce{C2H3+} ($m/z$ 27; green) and \ce{HNCCl+} ($m/z$ 62; black). The reduction of the \ce{CCl+} population (blue) is concurrent with the growth of \ce{C2H3+} (green) and \ce{HNCCl+} (black). Both of the primary product populations then reduce over time as the secondary product \ce{CH3CNH+} ($m/z$ 42; red) population grows from reactions with excess \ce{CH3CN}. \ce{CH3CNH+} is confirmed as a second order product because its maximum slope coincides with the maximum number of primary products. 
Experimental reaction rates are determined by fitting the reaction data to a pseudo-first order model. These curve fits are shown as lines in Fig. \ref{fig:reactioncurve}. Details of these fits are provided in the supplementary material. 

The primary product mass assignments, namely \ce{C2H3+} and \ce{HNCCl+}, given by the initial reaction of \ce{CCl+ + CH3CN} were verified by using different combinations of isotopologues. Specifically \ce{C^{37}Cl+} ($m/z$ 49) and \ce{CD3CN} ($m/z$ 44) were used to form four possible combinations of reactants. Reaction curves were measured for each of the four unique pairs and mass peak shifts were recorded for each case. Specifically, when the reaction proceeded with \ce{C^37Cl+ + CH3CN}, only one of the primary products shifted, $m/z~62\rightarrow64$ (\ce{HNC^37Cl+}), identifying it as the only chlorine-containing product. In the case of \ce{CCl+ + CD3CN}, both primary products shifted: $m/z~27\rightarrow30$ (\ce{C2D3+}), and $m/z~62\rightarrow63$ (\ce{DNCCl+}). Furthermore, the secondary product shifted, $m/z~42\rightarrow46$ (\ce{CD3CND+}). In the final case, \ce{C^37Cl+ + CD3CN}, the mass shifts were consistent with the aforementioned products. An additional process occurs in reactions involving \ce{CD3CN}, which produces a small amount of a tertiary product $m/z$ 45, assigned to \ce{CD3CNH+}. This tertiary process occurs possibly by either from H-D swapping or from contributions from a small number of contaminant ions remaining from the initial ion loading scheme (any given contaminant constitutes $\leq5\%$ of 150-250 initial \ce{CCl+} numbers). The isotopologue reaction curves are plotted in the supplementary material. Extrapolated rate constants and branching ratios from these reaction curves are provided in Tables \ref{tab:rates}-\ref{tab:rates2}.  

\begin{table}
    \centering
    \caption{Rate constants for isotopological variations of \ce{CCl^+ + CH3CN} primary products. `X' represents a hydrogen or deuterium from acetonitrile, and corresponds to the isotopologue used. Rates are in units of $\times 10^{-9}$\,cm$^3$/s, and reported statistical uncertainty is the calculated 90\% confidence interval.}
    \label{tab:rates}
    \begin{tabular}{lcccccc}
    \hline
    \hline
       Reactants  & & \ce{C2X3^+}& & \ce{XNCCl^+} && total  \\
    \hline
     \ce{CCl^+ +CH3CN} &&1.6 $\pm$ 0.5 && 2.2 $\pm$ 0.5 && 3.8 $\pm$ 0.4   \\
         \vspace{-3mm} 
         &&&\\
         \ce{C^37Cl^+ +CH3CN} &&  2.9 $\pm$ 0.7 && 3.0 $\pm$ 0.7 & &5.9 $\pm$ 0.3     \\
     \ce{CCl^+ +CD3CN} & &2.4 $\pm$ 0.5  & &3.0 $\pm$ 0.5 && 5.4 $\pm$ 0.3     \\ 
     \ce{C^37Cl^+ +CD3CN} & &2.9 $\pm$ 0.8  & &3.4 $\pm$ 0.8 & &6.3 $\pm$ 0.3     \\ 
     \hline\hline
    \end{tabular}
\end{table}

The measured rate constants for primary products of \ce{CCl+ + CH3CN} are reported in Table \ref{tab:rates}. The Langevin capture model is a natural starting place for the analysis of experimental reaction rate constants, as it is the simplest and most general approach for predicting rate constants in this regime. Notably temperature-independent, this theory estimates the likelihood of collisions between an ion and a neutral nonpolar molecule. The Langevin rate constant was found to be $k = 1.11\times10^{-9}$\,cm$^3$/s, 3-6 times smaller than the total reaction rate constant. This underestimation is most likely due to the polar nature of neutral \ce{CH3CN}, which is not accounted for in Langevin theory. Average dipole orientation (ADO) theory expands on Langevin theory to account for the polarity of the neutral reactant and should show closer agreement with the measured total reaction rate constant.\cite{SuBowers1973} This is reflected in the fact that \ce{CH3CN} has a rather large dipole-locking constant ($c$) of $\sim$0.25, leading to k$_{ADO,unsub} = 3.74\times10^{-9}$\,cm$^3$/s (calculated with the reduced mass of unsubstituted reactants). Our measured total reaction rate constant for \ce{CCl+ + CH3CN}, $3.8\pm0.7\times10^{-9}\,\text{cm}^3/\text{s}$ (see Table \ref{tab:rates}), reflects good agreement with ADO theory. This agreement testifies to the high degree of efficiency of the \ce{CCl+ + CH3CN} reaction, where effectively every ion-molecule collision results in the formation of new reaction products, with little reformation of the reactants (\textit{vide infra}). The high reactivity of \ce{CCl+} toward acetonitrile stands in stark contrast to much of the previous work on the reaction kinetics of this ion with neutral molecules.

The isotope substituted total reaction rate constants (also in Table \ref{tab:rates}) agree fairly well with the measured rate constant for \ce{CCl+ + CH3CN}, but do trend faster, between $5.4-6.4\times10^{-9}$cm$^3$/s, compared to the unsubstituted total reaction rate constant. This trend is not precisely captured by ADO theory, which predicts a very small ($\leq5$\%) reduction in the rate constant for both \ce{C^37Cl+} and \ce{CD3CN} substitutions. There is precedence for the trend of increased rate constant upon isotope substitution. Indeed, recently, this inverse kinetic isotope effect has been observed using a similar apparatus and Coulomb crystal environment by monitoring the charge exchange reaction between \ce{Xe+} and \ce{NH3} or \ce{ND3}. This effect, which was suggested to be due to intramolecular vibrational redistribution (IVR) occurring at a faster rate, and to a higher density of states in the deuterated ammonia.\cite{Petralia2020} It is possible that we are observing a similar effect here. It should be emphasized that we use a Bayard-Alpert hot cathode ionization gauge to measure the partial pressure of \ce{CH3CN} gas in the chamber. While sensitivity factors for the gases used in this study have been previously measured, they are not well characterized at pressures of $10^{-9}-10^{-10}$\,Torr (current regime). This systematic uncertainty is difficult to quantify, and is not reflected in our reported uncertainties. For this reason, we do not make a definitive assessment as to whether we are observing an inverse kinetic isotope effect. Instead, more significance is placed on the determination of branching ratios (see Table \ref{tab:branchingratio}) and assignments of chemical formulas and structures of observed reaction products, rather than to individual rate constant measurements. 

\begin{table}[h]
    \centering
    \caption{Branching ratios for primary products by isotopological variations of \ce{CCl^+} + \ce{CH3CN} reaction. The calculated branching ratio represents the fraction of protonated acetylene rate constant, divided by the total \ce{CCl+} decay rate constant. `X' represents a hydrogen or deuterium, and corresponds to neutral reactant.}
    \label{tab:branchingratio}
    \begin{tabular}{lc}
    \hline\hline
     &  Branching Ratio\\
       Reactants   & (k(\ce{C2X3+})/k$_{total}$)\\
    \hline
     \ce{CCl^+ +CH3CN} & 0.43 $\pm$ 0.16\\
    \ce{C^37Cl^+ +CH3CN} & 0.50 $\pm$ 0.17\\
     \ce{CCl^+ +CD3CN}  & 0.44 $\pm$ 0.11\\ 
      \ce{CCl^+ +CD3CN}  & 0.46 $\pm$ 0.17\\ 
      \hline\hline
    \end{tabular}
\end{table}
 
The branching ratios shown in Table \ref{tab:branchingratio} are nearly 50\% for each of the primary products; here reported as the rate of the \ce{C2H3+} production over the sum of both primary product rate constants. If all products branched from the same final step of the potential energy surface (see Fig. \ref{fig:CCl+CH3CN_PES}), the more exothermic product, \ce{HNCCl+}, might be expected to be favored. However, as will be discussed in section \ref{ModelPES}, the potential energy surface is much more complex, with the existence of branching pathways, as well as multiple isomers of products. This necessitates an energy grained master equation approach to obtain quantitative branching ratio predictions.

Secondary reactions with excess \ce{CH3CN} are comprised of a proton transfer from either \ce{C2H3+} or \ce{HNCCl+} forming \ce{CH3CNH+}. Analysis of the kinetics for these reactions is more straightforward, and the relative proton affinities of the neutral molecules guide our expectations for the stability of the products. \ce{CH3CN} has a larger proton affinity than either \ce{NCCl} or \ce{C2H2} (see supplementary material for calculated values), and thus both primary products transfer a proton to neutral \ce{CH3CN} to form the secondary product \ce{CH3CNH+}. Reaction dynamics predicted by relative proton affinities has precedence in ion-neutral gas-phase chemistry, and bounds on proton affinities have been determined by examining which proton transfers do or do not take place.\cite{Burt1970} In addition, these reactions are both energetically favorable, as per the reaction thermodynamics reported in Eqns. \ref{s3}-\ref{s4}. As for the relative rate constants calculated for the second order reactions, ADO theory predicts a slightly larger rate constant for the \ce{C2H3^+ +CH3CN} reaction ($4.3\times 10^{-9}$\,cm$^3$/s) due to its smaller reduced mass as compared to \ce{HNCCl^+ +CH3CN} ($3.5\times 10^{-9}$\,cm$^3$/s). This trend is consistent with the reported experimental reaction rate constants in Table \ref{tab:rates2}. Overall, there is reasonable agreement within the experimental uncertainty between the ADO calculated rate constants and those measured experimentally.

 \begin{table}[h]
    \centering
    \caption{Rate constants for isotope variations of \ce{CCl^+ + CH3CN} secondary products. `X' represents a hydrogen or deuterium from \ce{CH3CN}, and corresponds to the isotopologue used.  Rates are in units of $\times 10^{-9}$\,cm$^3$/s, and reported statistical uncertainty is the calculated 90\% confidence interval.}
    \label{tab:rates2}
    \begin{tabular}{lc}
    \hline
    \hline
       Reactants  & \ce{CX3CNX^+}  \\
    \hline
     \ce{C2H3^+ +CH3CN}  &   4.2 $\pm$ 1.7     \\
     \ce{HNCCl^+ +CH3CN}   &      4.1 $\pm$ 1.2  \\
     \\
    \ce{C2H3^+ +CH3CN}  &   6.2 $\pm$ 2.0    \\
    \ce{HNC^37Cl^+ +CH3CN} &  3.8 $\pm$ 1.1  \\
     \\
     \ce{C2D3^+ +CD3CN}     & 6.0 $\pm$ 1.5  \\ 
     \ce{DNCCl^+ +CD3CN}  &   4.4 $\pm$ 0.9  \\ 
 \\
     \ce{C2D3^+ +CD3CN}    & 6.2 $\pm$ 2.3  \\ 
     \ce{DNC^37Cl^+ +CD3CN}  &   5.9 $\pm$ 1.9  \\ 
     \hline\hline
    \end{tabular}
\end{table}

\subsection{Modelling the \ce{CCl+ + CH3CN} reaction}\label{ModelPES}
\begin{figure*}[!ht]
    \centering
    \includegraphics[width=17cm]{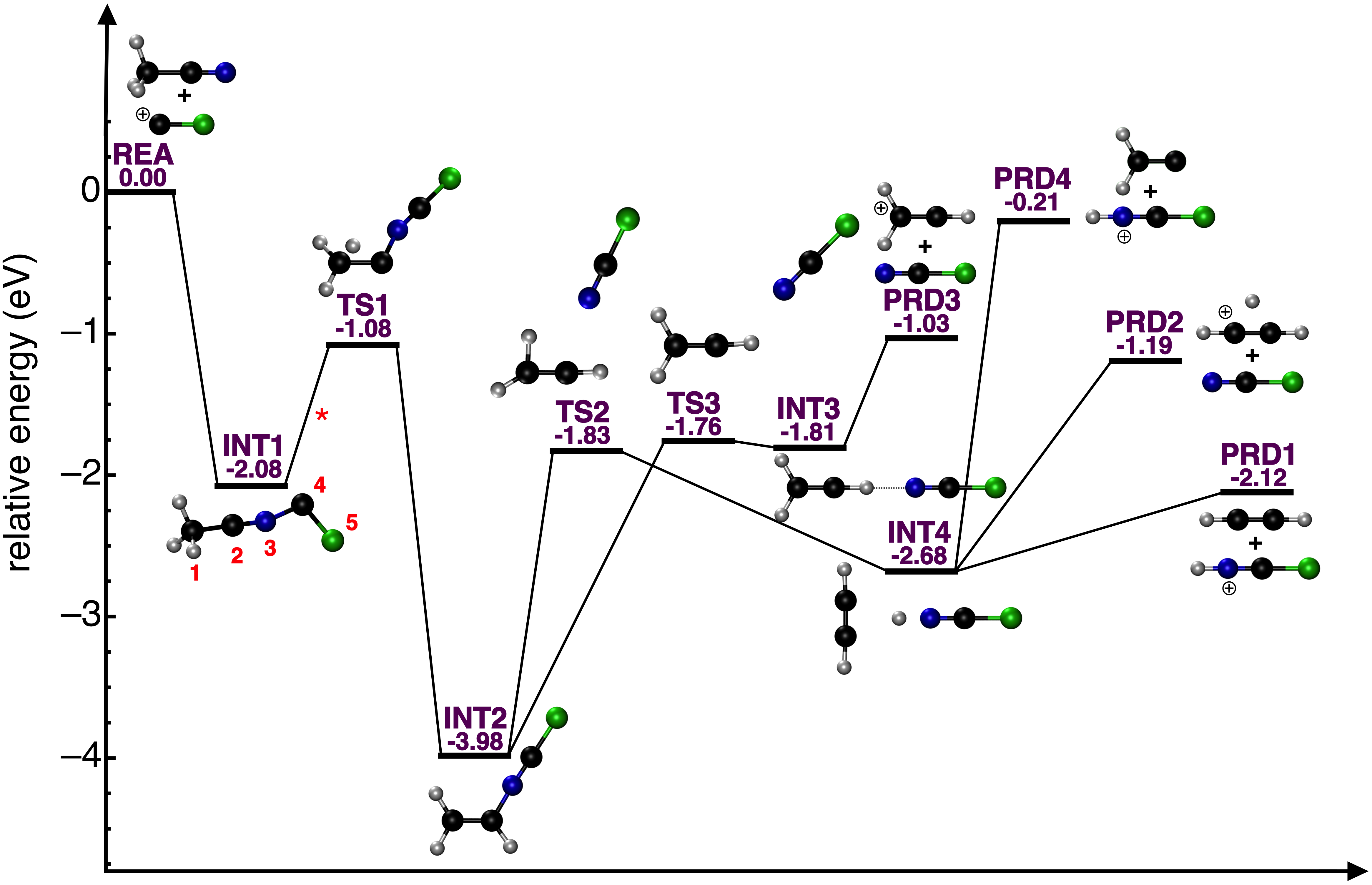}
    \caption[]{
        Potential energy surface for \ce{CCl^+ + CH3CN}, depicting equilibrium geometries connecting the reactants (REA) to the products (PRD1, PRD2, PRD3, and PRD4). In REA, PRD1, PRD2, PRD3, and PRD4, the bare `+' denotes infinite distance between the ion-neutral pair, while the {\tiny\circled{+}} symbol indicates the ion of the ion-neutral pair. Geometries were calculated at MP2/aug-cc-pVTZ level, with CCSD(T)/CBS//MP2/aug-cc-pVTZ energies. `INT' refers to intermediate states, while `TS' indicates transition states.
        Asterisk denotes a step with a very shallow well (depending on the level of theory), which is discussed in detail in the supplementary material. 
         }
    \label{fig:CCl+CH3CN_PES}
\end{figure*}

The potential energy surface shown in Fig. \ref{fig:CCl+CH3CN_PES} represents a few plausible reaction pathways of the \ce{CCl+ + CH3CN} reaction. It is a result of quantum chemical calculations and is comprised of equilibrium structures that bridge the reactants and the observed products. The experimental conditions are cold and very low pressure, which therefore means that there is no quenching of the internal energy of any of the intermediate low energy structures. Furthermore, the stationary points along this reaction pathway are all exothermic with respect to the reactants, such that the reaction complex can sample all these intermediary states until it leaves the surface irreversibly. It is useful to consider the potential energy surface not only because it is an accessible way to explore the pathways to eventual exothermic products presented, but also because it provides a basis for the quantitative master equation-based kinetic modeling presented below. For clarity, the non-hydrogen atoms will be numbered C1, C2, N3, C4, Cl5, as marked on INT1 in Fig. \ref{fig:CCl+CH3CN_PES}.

In the presented potential energy surface, \ce{CCl+} and \ce{CH3CN} initially form the adduct INT1 as a bond is formed between N3 and C4. This structure then undergoes various changes in its bond lengths and angles isomerizing into the lower energy INT2 structure. INT2 can isomerize into INT4, which can dissociate without a barrier into PRD1 (\ce{HNCCl+ +HC2H}), PRD2 (\ce{C2H3+ +NCCl}; where \ce{C2H3+} is the non-classical bridge structure), or PRD4 (\ce{HNCCl+ +H2C2}; where \ce{H2C2} is the vinylidene isomer of \ce{C2H2}). Determining the exact chemical identity of the \ce{C2H2} isomer is beyond the scope of this study: while the $m/z$ of ionic products is known based on the mass spectra, neutral products are speculative since they cannot be observed experimentally. 

INT2 can also isomerize to INT3, which leads to the barrierless dissociation into PRD3, the classical ``Y'' \ce{C2H3+} structure and \ce{NCCl}. The isomerization barrier between the two isomers of \ce{C2H3+} has been the subject of rigorous computational and experimental studies, and was found to be 4.8\,meV as calculated at the CBS-APNO level of theory.\cite{Psciuk2007,Crofton1989,Sharma2006} Regardless of which isomer is produced in this reaction, both isomers are energetically allowed, with exothermicity larger than the isomerization barrier. Therefore, either \ce{C2H3+} isomer may be the experimentally observed cation.

All of the outlined products are exothermic with respect to the reactants and there are only submerged barriers in the potential energy surface. This indicates that both products are likely to form, which is perhaps reflected in the experimentally observed branching ratios being equal. This observation is tested below through RRKM theory/master equation kinetic modeling.

To the best of our knowledge, there are no previous measurements for reactions of \ce{CCl+} with any nitriles with which to compare the current results. It does appear to be significant that the elucidated potential energy surface requires cleaving of the \ce{C#N} bond of \ce{CH3CN}. However, this is perhaps unsurprising given that once a bond is formed between the two reactants, more electron density will be pulled toward the more electronegative chlorine group. This is demonstrated in the first step of the PES, when INT1 (see Fig. \ref{fig:CCl+CH3CN_PES}) is formed. Two C-N bonds are of importance to this discussion: the C2-N3 bond, which originated from \ce{CH3CN}, and the C4-N3 bond, where the carbon from \ce{CCl+} attaches to the terminal nitrogen of \ce{CH3CN}. The shift of electron density from the C2-N3 bond to the C4-N3 and C4-Cl5 bonds occurs in this first steps of this potential energy surface. On this surface, the shift of electron density between stationary points INT1 and TS1 (Fig. \ref{fig:CCl+CH3CN_PES}) suggests the \ce{C#N} functional group pairs with \ce{Cl} over \ce{CH3}, stabilizing the complex with respect to the reactants. This is perhaps intuitive, as the highly electronegative \ce{Cl} atom pulls electron density towards itself, forming a strong bond, further assisted by the electron donating methyl group of \ce{CH3CN}. 

All products that are observed in this study are possibly a result of this shift and subsequent cleavage. Using the \ce{^{13}CH3 ^{13}CN} isotopologue as the neutral reactant could possibly provide more convincing experimental evidence of the \ce{C#N} bond cleaving mechanism, however, the cost of the reagent was prohibitive. While unsuccessful attempts were made to find a reaction pathway that did not cleave this \ce{C#N} bond, this did not constitute an exhaustive search of the PES. Regardless of whether a reaction pathway without cleavage of the \ce{C#N} bond exists, this theoretical mechanism is interesting in its own right. 

\begin{figure*}[!ht]
    \centering
    \includegraphics[width=17cm]{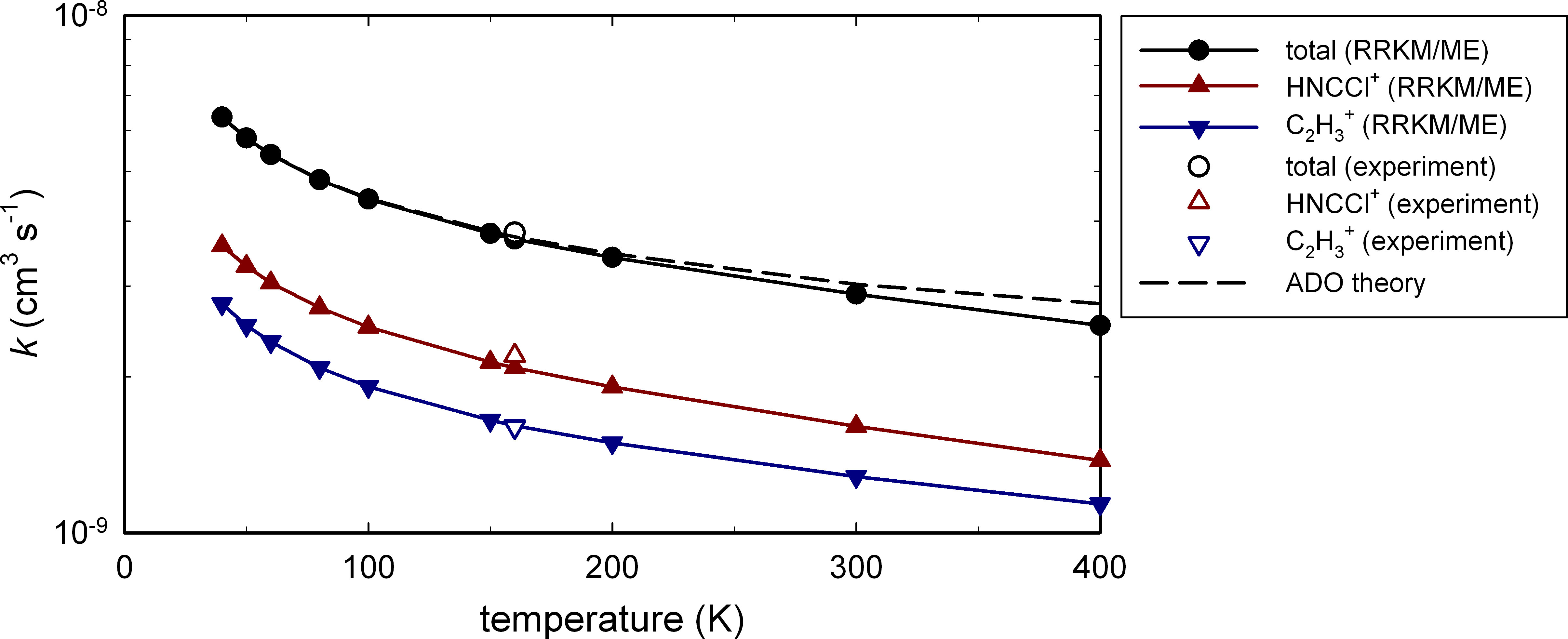}
    \caption[]{Theoretical (RRKM/ME) rate constants for the \ce{CCl+ + CH3CN} reaction as a function of temperature. Values are included for the overall reaction (total) and for the formation of product ions \ce{HNCCl+} (PRD1) and \ce{C2H3+} (PRD2 + PRD3). Included for comparison are the experimental measurements (at the effective temperature of 160\,K) and the ADO theory capture rate constants.}
    \label{fig:RRKM}
\end{figure*}

To gain further insight into the \ce{CCl+ + CH3CN} reaction, RRKM theory / master equation simulations were conducted on the basis of the potential energy surface reported in Fig. \ref{fig:CCl+CH3CN_PES} (with PRD4 excluded). Predicted rate constants are plotted in Fig. \ref{fig:RRKM} for the overall reaction and for formation of the PRD1 - PRD3 products as a function of temperature. Here, the overall rate constants reflect the ADO theory rates less any reverse dissociation of the ion-molecule complex back to the reactants. Also included in Fig. \ref{fig:RRKM} is the experimental measurement made here and the ADO theory capture rate constants.

Fig. \ref{fig:RRKM} indicates that the total rate constant is in good agreement with the experimental value, which in turn is similar to the ADO capture value. This reflects the high efficiency of the \ce{CCl+ + CH3CN} reaction, which leads almost exclusively to new products. This is in turn attributed to both the low barriers for \ce{CH3CNCCl+} isomerization and the availability of dissociation channels for the subsequent isomers at below the reactant energy. Only at temperatures of around 300 K and above is the reverse dissociation channel significant, resulting in the predicted rate coefficients to fall below the upper limit set by ADO theory.

Branching between the \ce{C2H3+} and \ce{HNCCl+} product ions is approximately 50:50, again in accord with the experiments. Interestingly, product PRD3 is predicted to be the dominant pathway to  \ce{C2H3+}, suggesting that it is formed in the classical, yet slightly higher-energy, vinylium form. This result is attributed to transition states TS2 and TS3 throttling the reaction flux from INT2 to a similar extent. Once TS2 is overcome, dissociation to PRD1 outcompetes all other channels (including PRD2), due to its low energy and high entropy. Following TS3, INT3 prefers to dissociate further to PRD3 than to isomerize back to INT2, presumably due to the loose forward dissociation being highly favored in terms of entropy.

\section{Conclusion and outlook}
The gas-phase reaction of \ce{CCl^+ + CH3CN} is presented, with primary products \ce{C2H3+} and \ce{HNCCl+} formed in approximately equal yields, and both channels producing a \ce{CH3CNH+} secondary product. The LIT TOF-MS used in this study enables experimental conditions of low pressures and collisional energies, limiting the reaction dynamics to exothermic pathways without quenching the internal energy of the reaction complex. In addition, the high mass resolution afforded by the TOF-MS yields methodical product identification that is supported by isotope substitution and quantum chemical calculations. The presented potential energy surface pathways indicate a series of equilibrium structures shifting electron density from the original \ce{CH3CN} \ce{C#N} bond to the new \ce{C#N} bond formed with the carbon of \ce{CCl+}. The experimental rate constants were reported and compared to Langevin and ADO theory capture rates, as well as to detailed master equation / RRKM theory-based simulations of the reaction kinetics on a multiple-channel multiple-well potential energy surface. ADO theory, which includes the polarity of the neutral reactant, is in good agreement with the observed experimental primary product rate constants. The master equation modeling indicates that reaction is highly efficient, with the total rate constant predicted to approach the capture rate constant. Moreover, these calculations reproduce the experimentally observed branching fractions between the primary ionic products \ce{C2H3+} and \ce{HNCCl+}. Although \ce{CCl+} has been predicted to not react with several neutrals, here, we see this is not the case, which is consistent with the previously observed reactions with \ce{C2H2}.\cite{Catani2020} This study presents the first example of this class of gas-phase reactions to be studied in a regime more closely comparable to that of the ISM (namely low pressure and temperature), and should aid in predicting the behavior of halogenated carbocations and nitriles in this region.

Future studies could further characterize \ce{CH3CN} with analogous reactions of various halogenated carbocations such as the astrochemically relevant ion \ce{CF+}. In theory, a reaction of \ce{CF+} with \ce{CH3CN} would behave similarly, and the even more electronegative fluorine might be expected to reproduce chlorine's behavior here. This would be particularly relevant to verify, as the presence \ce{CF+} in the ISM is more firmly established. It would also be interesting to study the effects of various functional groups (possibly more electron donating or withdrawing) attached to the \ce{C#N} in lieu of the methyl of \ce{CH3CN}. For example, benzonitrile \ce{C6H5(CN)} with its attached phenyl group could help stabilize intermediates or primary products and thus possibly shift the observed reaction rates.
Studying the reaction of \ce{CCl+} with various substituted nitriles might help elucidate a a trend in nitrile reactivity in this low pressure and cold regime. Overall, probing the relative \ce{C#N} bond strength across nitriles might contribute to the understanding and predictions of the formation and reactivity of the nitriles present throughout the ISM. Although further isotope tagging is necessary to absolutely verify the experimental reaction mechanism, the computational results are suggestive, and open questions for the role and reactivity of the \ce{C#N} bond in nitriles.

For the LIT-TOFMS apparatus, future directions also include the integration of a traveling wave Stark decelerator\cite{Shyur_2018a, Shyur_2018b} to expand control over the internal and external energies of polar neutral molecules. The ability to slow molecules down into the millikelvin regime allows the elucidation of whether quantum mechanical effects to play a greater role ion-neutral chemical dynamics. In this way, it presents an opportunity to both understand this class of reactions at a fundamental level, as well as further our understanding of ISM chemistry.

\section*{Supplementary Material}
See supplementary material for expanded experimental results, including: plots of averaged total ions over reaction times, details of reaction curve fits, and reaction data, as well as curves for isotopologue substituted reactions. See also for computational results in more detail: the full potential energy surface, geometries for stationary points at MP2/aug-cc-pVTZ level of theory, and geometries and energies for reaction limits at CCSD(T)/CBS//CCSD/aug-cc-pVTZ level of theory.

\begin{acknowledgments}
 This work was supported by the National Science Foundation (PHY-1734006, CHE-1900294) and the Air Force Office of Scientific Research  (FA9550-16-1-0117). GdS is supported by an Australian Research Council Future Fellowship (FT130101340). 
\end{acknowledgments}

\section*{Data Availability}

The data that support the findings of this study are available in the supplementary material and from the corresponding author upon reasonable request.

\section*{References}

\end{document}